\documentclass[aip,jcp,12pt,showkeys,reprint]{revtex4-1}
\usepackage{amsmath}
\usepackage{amsthm}
\usepackage{graphicx}
\usepackage{dcolumn}
\usepackage[colorlinks]{hyperref}
\allowdisplaybreaks

\begin{document}
\title{
Development of polaron-transformed explicitly correlated full configuration interaction method for investigation of quantum-confined Stark effect in GaAs quantum dots
}


\author{Christopher J. Blanton}
\affiliation
{Department of Chemistry, Syracuse University, Syracuse, New York 13244, USA}
\author{Christopher Brenon}
\affiliation{East Syracuse Minoa High School,Syracuse, NY 13057}
\author{Arindam Chakraborty}
\email[corresponding author: ]{archakra@syr.edu}
\affiliation
{Department of Chemistry, Syracuse University, Syracuse, New York 13244, USA}

\date{\today}

\begin{abstract}
The effect of external electric field on electron-hole 
correlation in GaAs quantum dots is investigated. The electron-hole Schr\"odinger equation in the presence of external electric field is solved using explicitly correlated full configuration interaction (XCFCI) method and accurate exciton binding energy and electron-hole recombination probability are obtained. The effect of the electric field was included in the 
1-particle single component basis functions by performing variational polaron 
transformation. The quality of the wavefunction at small inter-particle distances
was improved by using Gaussian-type geminal function that depended explicitly on 
the electron-hole separation distance. The parameters of the explicitly correlated function
were determined variationally at each field strength. The scaling of total exciton energy,
 exciton binding energy, and electron-hole recombination probability with respect to
 the strength of the electric field was investigated. It was found that a 500 kV/cm change in electric field reduces the binding energy and recombination probability 
by a factor of 2.6 and 166, respectively. The results show that
the eh-recombination probability is affected much more strongly by the electric field than the exciton binding energy. Analysis using the polaron-transformed basis indicate that the exciton binding should asymptotically vanish in the limit of large field strength.
\end{abstract}

\pacs{31.10.+z, 31.15.-p}                      
\keywords{ quantum-confined Stark effect,
semiconducting quantum dots, 
exciton binding energy, external electric fields, polaron transformation,
explicitly correlated, variational method,
 gallium arsenide} 

\maketitle

\section{Introduction}
The  influence of external electric field on optical properties
of semiconductors has been studied extensively using 
both experimental and theoretical techniques.
In bulk semiconductors the shift in the optical 
absorbing due to the external field is known as the 
Franz-Keldysh effect.\cite{seeger2004semiconductor}
In quantum wells and quantum dots, application of electric field  
has shown to modify the optical properties of nanosystems 
and is known as the 
quantum-confined Stark effect (QCSE).\cite{miller1984band,miller1985electric}
The application of the external field induces various modifications 
in the optical properties of the nanomaterial including, 
absorption coefficient, spectral weight of transitions, 
and change in $\lambda_\mathrm{max}$ of the absorption spectra. In certain cases, 
the applied field can lead to exciton ionisation.\cite{perebeinos2007exciton} 
The quantum-confined Stark effect has found application in the 
field of electro-absorption modulators,\cite{bimberg2012quantum} 
solar cells\cite{yaacobi2012combining} and the light-emitting 
devices.\cite{de2012quantum}
Recent experiments by Weiss et al. on semiconductor quantum dots have shown that 
the QCSE can also be enhanced by the presence of heterojunctions.\cite{park2012single}
In some cases, the QCSE can be induced chemically because of close 
proximity to ligands.\cite{yaacobi2012combining} 
The QCSE also plays a major role in electric field dependent photoconductivity
in CdS nanowires and nanobelts.\cite{li2012electric} 
Electric field has emerged as one of the tools to control and customize quantum dots
as novel light sources. In a recent study, electric field was 
used in generation and control of polarization-entangled photons using GaAs quantum dots.\cite{ghali2012generation}
It has been shown that the coupling between stacked quantum dots can be modified 
using electric field.\cite{talalaev2006tuning} 
\par
The QCSE
has been investigated using various theoretical techniques including
perturbation theory,\cite{Jaziri1994171,Kowalik2005,Xie20091625,He2010266,Lu2011,Chen2012786}
variational techniques,\cite{Kuo200011051,Dane2008278,
Barseghyan2009521,Duque2010309,Dane20101901,Kirak2011,
Acosta20121936} and configuration interaction method. \cite{Bester2005,Szafran2008,Reimer2008,
Korkusinski2009,Kwaniowski2009821,Pasek2012,Luo2012,
Braskan2001775,Braskan20007652,Corni2003853141,Lehtonen20084535}
In the present work, development of explicitly correlated 
full configuration interaction (XCFCI) method is presented
for investigating effect of external electric field on quantum dots and wells.
The XCFCI method is a variational method in which the conventional CI 
wavefunction is augmented by explicitly correlated Gaussian-type 
geminal functions.\cite{JoakimPersson19965915} 
The inclusion of explicitly correlated function in the 
form of the wavefunction is important for the following two reasons.
First, the addition of the geminal function increases the 
convergence of the FCI energy with respect to the size of the 
underlying 1-particle basis set.\cite{Prendergast20011626}
Second, inclusion of explicitly
correlated function 
improves the form of the electron-hole wavefunction at 
small inter-particle distances which is important 
for accurate calculation of electron-hole recombination
probability.\cite{RefWorks:2334,Wimmer2006,RefWorks:4030}
The effect of explicitly correlated function 
on the convergence of CI energy has been investigated by 
Prendergast et al.\cite{Prendergast20011626} and
is directly related to 
accurate treatment of the Coulomb singularity in the Hamiltonian.\cite{Hattig20124,Kong201275,Prendergast20011626}
Varganov et al. have demonstrated the applicability of 
geminal augmented multiconfiguration self-consistent field
wavefunction for many-electron systems.\cite{varganov2010variational}
 Elward et al. have 
also performed variational calculation using
explicitly correlated wavefunction for treating 
electron-hole correlated in quantum dots.\cite{RefWorks:4030,RefWorks:4031} 
\par
One of the important features of the XCFCI method presented here
is the inclusion of the external field in the ansatz of the wavefunction.
This is achieved by defining
a new set of  field-dependent coordinates which are generated by performing 
variational polaron transformation\cite{harris1985variational} and 
recasting the original Hamiltonian in terms of the field-dependent
coordinates. 
The variational polaron transformation was introduced by  Harris 
and Silbey for studying quantum dissipation phenomenon in the spin-boson system\cite{harris1985variational} and is used in the present work
because of the mathematical similarity between the spin-boson and the 
field-dependent electron-hole Hamiltonian.  
\par
The remainder of this article is organized as follows. The important features of
the XCFCI method are summarized in Sec. \ref{sec:xcfci}, construction of 
the field dependent basis functions is presented in Sec. \ref{sec:polaron},
the application of the XCFCI method using field-dependent basis is presented 
in Sec. \ref{sec:results}, and the conclusion are provided
in Sec. \ref{sec:conclusion}.

\section{Theory}
\subsection{Explicitly correlated full configuration interaction}
\label{sec:xcfci}
The field dependent electron-hole Hamiltonian is defined as\cite{RefWorks:4060,RefWorks:2174}
\begin{align}
\label{eq:ham}
	H &= 
	-\frac{\hbar^2}{2m_{\mathrm{e}}}\nabla^2_{\mathrm{e}}
	-\frac{\hbar^2}	{2m_{\mathrm{h}}}\nabla^2_{\mathrm{h}} 
	+ v^\mathrm{ext}_\mathrm{e} 
	+ v^\mathrm{ext}_\mathrm{h} \\ \notag          
	&- \frac{1}{\epsilon  \vert \mathbf{r}_{\mathrm{eh} } \vert} 
	+ \vert e\vert \mathbf{F} \cdot (\mathbf{r}_{\mathrm{e}}-\mathbf{r}_{\mathrm{h}})
\end{align}
where $m_{\mathrm{e}}$ is the mass of the electron,
$m_{\mathrm{h}}$ is the mass of the hole,
$\epsilon$ is the dielectric constant,
and 
$\mathbf{F}$ is the external electric field.
The external potential $v^\mathrm{ext}_\mathrm{e}$ 
and $v^\mathrm{ext}_\mathrm{h}$ represent the confining potential 
experienced by the quasi-particles.  
The form of the XCFCI wavefunction 
is defined as
\begin{align}
\label{eq:xcfci}
	\Psi_\mathrm{XCFCI}
	&= \hat{G}
	\sum_k
	c_k \Phi_k
\end{align} 
where $c_k$ is the CI coefficient and $\Phi_k$ are basis functions. The operator $\hat{G}$
is known as the geminal operator and is an explicit function of $r_\mathrm{eh}$
and is defined as
\begin{align}
	\hat{G} 
	= \sum_{i=1}^{N_\mathrm{e}}
	  \sum_{j=1}^{N_\mathrm{h}} 
	  \sum_{k=1}^{N_\mathrm{g}} 
	  b_{k}e^{-\gamma_k r_{ij}^2},
\end{align}
where $N_\mathrm{g}$ is the number of Gaussian functions included in the 
expansion, $N_\mathrm{e}$ and $N_\mathrm{e}$ are the number of electrons
and holes, respectively. The parameters $b_k$ and $\gamma_k$ used in the 
definition of the geminal operator are obtained variationally. 
The construction of the basis functions
used in the definition of XCFCI wavefunction in Eq. \eqref{eq:xcfci} will 
be discussed in Sec. \ref{sec:polaron}.
The XCFCI calculation is performed in two steps. In the first step, the 
parameters of geminal operator are obtained variationally by performing
the following minimization
\begin{align}
	E[G_{\mathrm{min}}]
	&= 
	\min_{b_k,\gamma_k}
	\frac{\langle G \Phi_0 \vert H \vert G \Phi_0 \rangle }
	{\langle G \Phi_0 \vert G \Phi_0 \rangle} . 
\end{align} 
In the second step, the expansion coefficients $\{c_k\}$ are obtained variationally
and are defined by the following minimization procedure
\begin{align}
\label{eq:Excfci}
	E_\mathrm{XCFCI} 
	&=
	\min_{\mathbf{c}} 
	\frac{\langle \Psi_\mathrm{XCFCI} \vert H \vert \Psi_\mathrm{XCFCI}  \rangle }
	{\langle \Psi_\mathrm{XCFCI} \vert \Psi_\mathrm{XCFCI}  \rangle } .
\end{align}
The above equation can be rewritten as a FCI calculation of transformed operators
\begin{align}
	E_\mathrm{XCFCI} 
	&=
	\min_{\mathbf{c}} 
	\frac{\langle \Psi_\mathrm{FCI} \vert \tilde{H} \vert \Psi_\mathrm{FCI}  \rangle }
	{\langle \Psi_\mathrm{FCI} \vert \tilde{1} \vert \Psi_\mathrm{FCI}  \rangle }, 
\end{align}
where the transformed operators are defined as
\begin{align}
\label{eq:htilde}
	\tilde{H} &= G_\mathrm{min}^\dagger H G_\mathrm{min}, \\
\label{eq:stilde}
	\tilde{1} &= G_\mathrm{min}^\dagger G_\mathrm{min}.
\end{align}
The exact expression of the transformed operators in Eq. \eqref{eq:htilde}
and \eqref{eq:stilde} and discussion relevant to their derivation has 
been presented earlier in Ref. \onlinecite{RefWorks:4030,RefWorks:4031} and is 
not repeated here. The $E_\mathrm{XCFCI}$ reduces to conventional FCI energy in the limit of geminal function 
equals to 1
\begin{align}
	E_\mathrm{FCI} = \lim_{G \rightarrow 1} E_\mathrm{XCFCI}
\end{align}
We expect the $E_\mathrm{XCFCI}$ energy to be lower than the FCI energy for identical set of 
basis functions and earlier studies have shown this to be true.\cite{RefWorks:4031}
\par
After the successful completion of the XCFCI calculations, the field dependent exciton binding was calculated from the difference between the non-interacting and interacting 
ground state energies. Defining the non-interacting Hamiltonian as
\begin{align}
\label{eq:h0}
	H_0 &= \lim_{\epsilon^{-1} \rightarrow 0} H ,
\end{align}
the exciton binding energy is computed as
\begin{align}
\label{eq:Eb}
	E_\mathrm{B}[\mathbf{F}] 
	&=
	E_\mathrm{XCFCI} - E_0^{(0)},
\end{align}
where $E_0^{(0)}$ is defined in Eq. \eqref{eq:e0}
\begin{align}
\label{eq:e0}
	E_0^{(0)}
	&=
	\min_{\Psi}
	\frac{\langle \Psi \vert H_0 \vert \Psi \rangle}
	{\langle \Psi \vert \Psi \rangle}.
\end{align}
The field dependent electron-hole recombination probability is obtained from the 
XCFCI wavefunction using the following expression\cite{RefWorks:4030,RefWorks:4031}
\begin{align}
\label{eq:recomb}
	P_{\mathrm{eh}} [\mathbf{F}]
	= 
	\frac{\langle\Psi_\mathrm{XCFCI}
	\vert \delta(\mathbf{r}_{\mathrm{e}}-\mathbf{r}_{\mathrm{h}}) \vert
	\Psi_\mathrm{XCFCI} \rangle}
	{\langle\Psi_\mathrm{XCFCI} 
	\vert \Psi_\mathrm{XCFCI} \rangle}.
\end{align}
The exciton binding energy and the recombination probability are
functionals of the applied external field and are indicated explicitly in
Eq. \eqref{eq:Eb} and \eqref{eq:recomb}, respectively.

\subsection{Construction of field dependent basis set}
\label{sec:polaron}
One of the key features of the electron-hole Hamiltonian used in the present work
is the presence of the field-dependent term in Eq. \eqref{eq:ham}. Since the 
convergence of the CI expansion depends on the quality of the 
underlying 1-particle basis, it is desirable to construct and use efficient single
particle basis sets. In the present work, we have developed field-dependent
basis functions and the details of the derivation are presented as following. Starting with the expression of $H_0$ in Eq. \eqref{eq:h0}, 
the zeroth-order Hamiltonian is expressed 
as a sum of non-interacting electronic and hole Hamiltonians
\begin{align}
	H_0 = H_0^\mathrm{e} + H_0^\mathrm{h},
\end{align}
where the expression for the single-component non-interacting Hamiltonian is given as
\begin{align}
	H_0^\mathrm{e} 
	&=
	T_\mathrm{e} + v_\mathrm{e}^\mathrm{ext} 
	+ \vert e \vert \mathbf{F} \cdot \mathbf{r}_\mathrm{e} \\
	H_0^\mathrm{h} 
	&=
	T_\mathrm{h} + v_\mathrm{h}^\mathrm{ext} 
	- \vert e \vert \mathbf{F} \cdot \mathbf{r}_\mathrm{h}.
\end{align}
As seen from the above equation, the coupling between the external field and the 
quasiparticle coordinates is linear. The above Hamiltonian shares mathematical 
similarity with the spin-boson Hamiltonian that has been used extensively in 
quantum dissipative systems.\cite{weiss2008quantum}
In the present method, we perform analogous transformation which is defined by the 
follow equations
\begin{align}
\label{eq:polar}
	\mathbf{q}_\mathrm{e} &= \mathbf{r}_\mathrm{e} + \lambda_\mathrm{e} \mathbf{F} \\
	\mathbf{q}_\mathrm{h} &= \mathbf{r}_\mathrm{h} - \lambda_\mathrm{h} \mathbf{F}.
\end{align}
Similar to the polaron transformation in the spin-boson system, the 
coordinates of the quasiparticle experience a shift due to the presence of the
external field. \cite{weiss2008quantum}
 Using the method of variational polaron transformation by Harris
and  Silbey,\cite{harris1985variational}
the shift parameter $\lambda$  is determined variationally. 
The field-dependent electronic basis functions are obtained by 
first constructing the Hamiltonian matrix using Gaussian-type orbitals (GTO)
and then diagonalizing the resulting matrix
\begin{align}
	H_0^\mathrm{e} \Phi_i^\mathrm{e} 
	&=
	\epsilon_i^\mathrm{e} (\lambda_\mathrm{e}) \Phi_i^\mathrm{e}
	\quad i=1,\dots,M_\mathrm{e} \\
	H_0^\mathrm{h} \Phi_j^\mathrm{h} 
	&=
	\epsilon_j^\mathrm{h} (\lambda_\mathrm{h}) \Phi_j^\mathrm{h}
	\quad j=1,\dots,M_\mathrm{h}.
\end{align}
The value of the shift parameter is obtained variationally by 
minimizing the trace
\begin{align}
	\min_{\lambda} \sum_{i}^{M_\mathrm{e}} \epsilon_i^\mathrm{e}
	\implies \lambda_\mathrm{e}.
\end{align}
The $\lambda_\mathrm{h}$ is also obtained by a similar procedure. 
The electron-hole basis functions for the FCI calculations are constructed by taking a
direct product between the set of electronic and hole single-component basis sets
\begin{align}
	\{ \Phi_k \} 
	&=
	\{ \Phi_i^\mathrm{e} \}  \otimes \{ \Phi_j^\mathrm{h} \}.
\end{align}
The procedure described above 
is a general method that is independent of the exact form of the 
external potential. However if the external potential is of quadratic form, 
the field dependent zeroth-order single-component Hamiltonian has
an uncomplicated mathematical form and additional simplification can be achieved. 

\section{Results and Discussion}
\label{sec:results}
\begin{figure}[h!]
    \includegraphics[width=85mm]{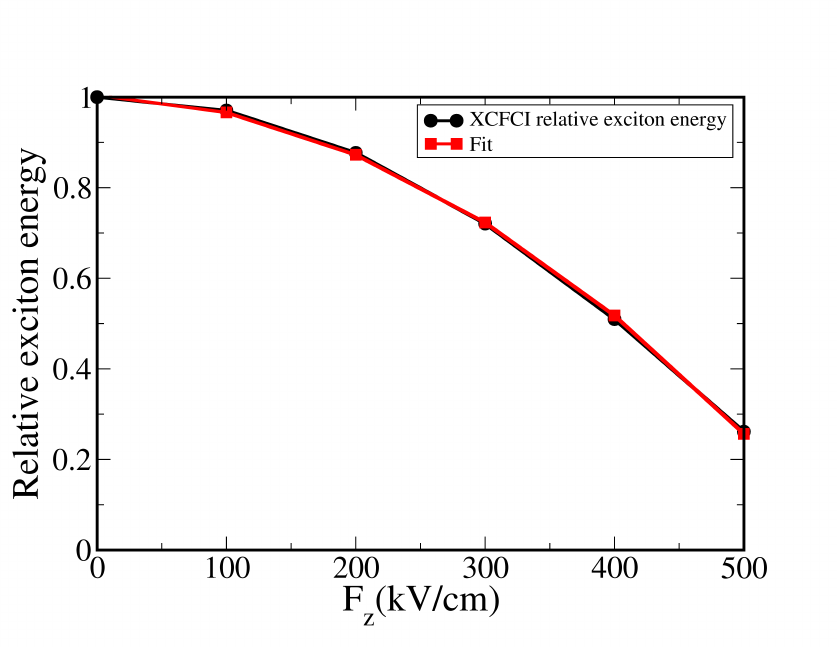}
    \caption{Relative exciton energy compared to the fit
     $E = (-2.7925\times 10^{-6})F_z^2 + (-7.0938\times 10^{-5})F_z + 1$.}
  \label{fig:toteng}
\end{figure}

\begin{figure}[h!]
  \begin{center}
    \includegraphics[width=85mm]{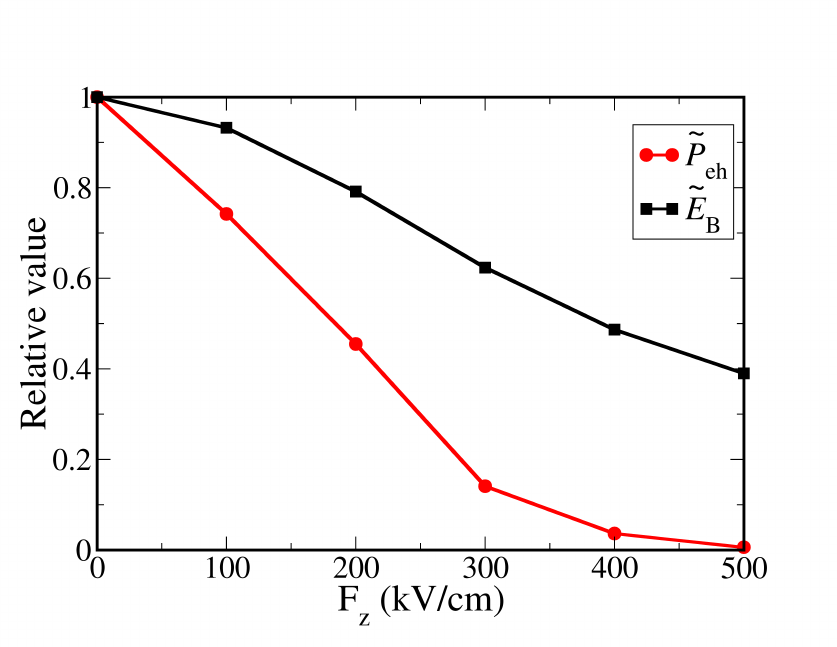}    
    \caption{Comparison of $\tilde{E}_{\mathrm{B}}$ and $\tilde{P}_{\mathrm{eh}}$ as a function of electric field strength.}
    \label{fig:relative}
  \end{center}
\end{figure}

The electron-hole Hamiltonian in Eq. \eqref{eq:ham} has 
been used extensively for studying optical rectification
\cite{RefWorks:4060,RefWorks:2174,
He2010266,RefWorks:4137,Chen2012786}
effect in 
GaAs quantum dots and all the system specific parameters 
were obtained from previous calculations on the GaAs system.\cite{RefWorks:4060,RefWorks:2174}
The parabolic confinement potential has found widespread 
applications\cite{Peeters19901486,Que199211036,Halonen19925980,
RefWorks:4035,Jaziri1994171,Rinaldi1996342,
RefWorks:4130,RefWorks:2155,Barseghyan2009521,
Taut2009,He2010266,Stobbe2010,RefWorks:4034,
RefWorks:4033,Kirak2011,Trojnar2011}  
in the study of quantum dots and was used in the present work 
to approximate the external potential term in the Hamiltonian.
All the parameters that are needed for complete description of the 
electron-hole Hamiltonian used in the calculations are presented 
in Table~\ref{tab:param}.
\begin{table}[h!]
  \caption{System dependent parameters used in the electron-hole Hamiltonian
   for the GaAs quantum dot \cite{RefWorks:4060,RefWorks:2174} }
    \begin{tabular}{cc}
      \hline
      Parameter & Value \\ 
      \hline
      $m_{\mathrm{e}}$ & $0.067m_{0}$ \\ 
      $m_{\mathrm{h}}$ & $0.090m_{0}$ \\ 
      $k_{\mathrm{e}}$ & $9.048\times 10^{-7}$ a.u. \\ 
      $k_{\mathrm{h}}$ & $1.122\times 10^{-6}$ a.u. \\ 
      $\epsilon$ & $13.1\epsilon_{0}$ \\ 
      \hline
    \end{tabular}
    \label{tab:param}
\end{table}
Following earlier work on the effect of
electric field on non-linear optical properties of GaAs quantum
dots,\cite{RefWorks:4060,RefWorks:2174} 
the external electric field was aligned along the z-axis and the field 
strength was varied from zero to 500 kV/cm.
Similar to the spin-boson Hamiltonian, 
the polaron transform resulted in shifted harmonic oscillators.\cite{weiss2008quantum} 
The eigenvalues and 
and eigenfunctions of the $H_0$ were obtained analytically, and the lowest ten 
eigenstates of the shifted harmonic oscillator Hamiltonian
were used in the construction of the 1-particle basis. The direct product 
between the electronic and the hole basis sets was performed to generate the 
electron-hole basis for the FCI calculations. The geminal minimization was
performed using  a set of three $\{b_k,\gamma_k\}$ parameters at each field strength,
and the optimized values are presented in Table \ref{tab:geminals}.
  \begingroup
  \squeezetable
  \begin{table*}
      \caption{Optimized geminal parameters used in the calculations of energy and recombination probability.}
      \begin{tabular}{c@{\hskip 10mm}c@{\hskip 10mm}c@{\hskip 10mm}c@{\hskip 10mm}c@{\hskip 10mm}c@{\hskip 10mm}c}
        \hline
        $F_z$ (kV/cm) & 0 & 100 & 200 & 300 & 400 & 500 \\ 
        \hline
        $b_1$ & 1.00 & 1.00 & 1.00 & 1.00 & 1.00 & 1.00 \\ 
        $\gamma_1$ & 0.00 & 0.00 & 0.00 & 0.00 & 0.00 & 0.00 \\ 
        $b_2$ & $1.40\times 10^{-1}$ & $9.99\times 10^{-1}$ & $4.99\times 10^{-2}$ & $5.78\times 10^{-3}$ & $1.00\times 10^{-2}$ & $5.59\times 10^{-3}$ \\ 
        $\gamma_2$ & $2.29\times 10^{-4}$ & $4.60\times 10^{-6}$ & $1.11\times 10^{-2}$ & $1.11$ & $1.00$ & $1.11$ \\ 
        $b_3$ & $4.35\times 10^{-2}$ & $1.08\times 10^{-1}$ & $8.90\times 10^{-2}$ & $1.67\times 10^{-2}$ & $2.00\times 10^{-2}$ & $1.58\times 10^{-2}$ \\ 
        $\gamma_3$ & $1.13\times 10^{-2}$ & $1.00\times 10^{-2}$ & $1.01\times 10^{-3}$ & $1.11\times 10^{-1}$ & $1.01\times 10^{-1}$ & $1.02\times 10^{-1}$ \\ 
        \hline
      \end{tabular}
      \label{tab:geminals}
  \end{table*}
  \endgroup  
The total exciton energy for the field-free case was found to be 
269.45 meV. The total exciton energy of the system as a function of the field strength is presented in Fig.~\ref{fig:toteng}.
It is seen that the total energy decreases with increasing field strength. Earlier studies on this system indicate that the exciton energy is a quadratic function of the applied field.\cite{Weiner1987842,Robinson20051}
To investigate the scaling of the total 
exciton energy with respect to the field strength, we have performed least-square fit of the calculated values with a second order polynomial and the results are presented in Fig. \ref{fig:toteng}.
The results from these calculations confirm that the quadratic scaling of the exciton energy
as a function of the field strength. 
The exciton binding energy was calculated using Eq. \eqref{eq:Eb} and was found to be
28.52 meV for the field-free case. 
The effect of the external field on the exciton binding energy was 
investigated by calculating the relative binding energy which is 
defined by the following equation
\begin{align}
	\tilde{E}_\mathrm{B}
	&=
	\frac{E_\mathrm{B}[\mathbf{F}]}{E_\mathrm{B}[\mathbf{F}=0]} .
\end{align}
It is seen from Fig. \ref{fig:relative} that the exciton binding energy decreases 
with increasing field strength.
As the field strength is increased from 0 to 500 kV/cm,
the exciton binding energy decreases by a factor of 2.6. In addition to calculation of binding energy, the effect of the field on electron-hole recombination probability 
was also investigated. Analogous to the relative binding energy, the relative 
recombination probability is defined as 
\begin{align}
	\tilde{P}_\mathrm{eh} &=
	\frac{P_\mathrm{eh}[\mathbf{F}]}{P_\mathrm{eh}[\mathbf{F}=0]},
\end{align}
and is presented in the Fig. \ref{fig:relative}. 
It is seen that the there is a
sharp decrease in the recombination probability with increasing field strength and
the recombination probability at 500 kV/cm is lower than the field-free case by a factor of 166.
One of the key results from this study is that exciton binding energy and eh-recombination probability are affected differently by the external electric field.  It is seen that the exciton binding energy and
eh-recombination probability follow different scaling with respect to field strength.

The polaron transformation also provides insight into the effect
electric field on the exciton binding energy 
in the limit of high field strengths. Starting with the transformation 
defined in Eq. \eqref{eq:polar}, the  electron-hole 
Coulomb interaction in the transformed coordinate can be expressed as
\begin{align}
	\frac{1}{\vert \mathbf{r}_\mathrm{e} - \mathbf{r}_\mathrm{h} \vert}
	&=
	\frac{1}{\vert (\mathbf{q}_\mathrm{e} - \mathbf{q}_\mathrm{h}) - (\lambda_e + \lambda_h) \mathbf{F} \vert } = v_\mathrm{eh}(\mathbf{q}) .
\end{align}
It is seen in the above equation that the above expression will be dominated by the 
field-dependent term in the limit of high field strength. A direct consequence of the 
above condition is that in the limit of high field strengths, we expect the 
exciton binding energy to be small
\begin{align}
\label{eq:limit}
	H(\mathbf{q}) \approx H_0(\mathbf{q}) 
	\implies E_\mathrm{B} \approx 0 
    \quad \quad  \mathrm{for}\, 1 \ll \vert \mathbf{F} \vert < \infty.
\end{align}
It is important to note that the above conclusion is independent of the choice of the
external potential. 

\section{Conclusion}
\label{sec:conclusion}
The effect of external electric field on exciton binding energy 
and electron-hole recombination probability was computed
using explicitly correlated full configuration interaction method. 
Field-dependent basis functions were used in the calculations
and a variational polaron transformation scheme was developed for 
construction of field-dependent basis functions. 
It was found that both exciton binding energy and 
electron-hole recombination probability decrease with increasing 
field strength. One interesting conclusion from this study is
that the binding energy and recombination probability follow 
different scaling with respect to the external electric 
field. For the range of field strengths studied, 
the recombination probability and exciton binding energy
decrease by a factor of 166 and 2.6, respectively. 
These results give important insight into the 
application of electric field for manipulating 
excitons in quantum dots.   

\begin{acknowledgments}
Acknowledgment is made to the donors of The American Chemical Society Petroleum
Research Fund (52659-DNI6) and to Syracuse University 
for support of this research.
\end{acknowledgments}

\newpage

\bibliographystyle{unsrt}
\bibliography{ref}

\end{document}